\documentstyle[12pt]{article}

\begin{document}

\hoffset = -1truecm
\voffset = -2truecm

\title{\bf
 Gravity and Domain Wall Problem
}

\author{
{\bf
 Balram Rai\thanks{E-mail: raib@itsictp.bitnet}
}\\
{\bf Goran Senjanovi\'{c}\thanks{E-mail: goran@itsictp.bitnet,
vxicp1::gorans.}}\\
\normalsize International Centre for Theoretical Physics, Trieste 34100,
{\bf Italy} }

\date{December 1992}
\newpage

\maketitle

\baselineskip = 16pt

\parskip = 18pt

\begin{abstract}
It is well known that the spontaneous breaking of discrete
 symmetries may lead to conflict with big-bang cosmology.
This is due to formation of domain walls which give
unacceptable contribution to the energy density of the universe.
On the other hand it is expected that gravity breaks global
symmetries explicitly. In this work we propose that this could
provide a natural solution to the domain-wall problem.

\end{abstract}

\hfull{ICTP Preprint: IC-92-414}

\newpage


{\bf Introduction}

  The idea that the discrete symmetries, especially the
 fundamental ones such as time-reversal and parity
 invariance, could be broken spontaneously
is old and appealing. Twenty years ago, in a pioneering
 work, Lee [1] suggested that CP (or T reversal ) may be
broken spontaneously at the cost of adding another Higgs
doublet to the standard model. It was shown later that
parity can also be broken spontaneusly [2]. Yet another
interesting example is a discrete symmetry needed in the
two Higgs doublet model to ensure natural flavour
conservation [3].
However, in a beautiful
 paper Zeldovich et.al. [4] investigated the cosmological
consequences of spontaneous breaking of a discrete symmetry
with conclusion that this would be in conflict with
cosmology. Kibble [5] and other authors [6] concluded the
same, although with a slightly different analysis. Since
then the
particle physics models with spontaneously broken discrete
symmetry have been considered unacceptable ( see below for some
possible exceptions ). In this letter we propose that
 the possibility of gravity leading to
violation of global (discrete) symmetries may provide an
attractive way out of this problem. We find it only natural that
the space-time dynamical effects of gravity would play
this role for the space-time discrete symmetries. Although the
arguments for violation of global discrete symmetry,
as we will describe in the following,
are speculative, the point we wish to emphasise is that even
expectedly tiny effects of gravity may suffice. In what
follows we first describe the above mentioned problem and
then discuss how gravity may possibly provide a solution.

{\bf The Problem}

      The spontaneous breaking of a discrete symmetry
 leads to the existence of domain walls, i.e. kink-like
classical solutions separating different degenerate vacua.
This can be illustrated by a simple example of a real
scalar field  $\phi$ with a
symmetry D: $\phi \rightarrow -\phi$  and a Lagrangian

\begin{equation}
{\cal L} = {1\over 2}{\partial}_{\mu}{\phi}{\partial}^{\mu}{\phi}
- {\lambda \over 4} ({\phi}^2 - v^2)^2
\end{equation}

\noindent where ${\lambda}$ is taken to be
positive in order to ensure that the
energy is bounded from below. The symmetry
D is spontaneously
broken, since the minimum of the potential is
given by ${\phi_{vac} = \pm v}$.
It is easy to show that this theory has a static domain-wall
like classical solution, say for a wall lying in x-y plane,

\begin{equation}
\phi_{cl} (z) = v tanh ({\surd}{\lambda}vz)
\end{equation}

\noindent which clearly connects vacua $-v$  and $+v$ as z
traverses from  -${\infty}$ to ${\infty}$. The field is
 different from its
vacuum values in a region of width
 ${\delta}_W \approx ({\surd}\lambda v)^{-1}$
 , determined by the scale
 of symmetry
breaking $v$. The scale of symmetry breaking also
determines
the energy density per unit area $\sigma \approx v^3$.

     Now, at high temperatures
the potential $V({\phi})$ receives an additional
contribution

\begin{equation}
\delta V = {\lambda \over 12}T^2{\phi}^2
\end{equation}

Since ${\delta}V$ is necessarily positive, for sufficiently high
temperature $T > T_c {\approx} v $ the symmetry is restored [9].
In the standard big-bang cosmological scenario, the field
 ${\phi}$
is expected to undergo a phase transition as the universe
cools down from $T > T_c$ to $T < T_c$ .
For separations larger than the correlation length or
horizon size around the time of phase transition, the field
${\phi}$ will independently take either of its vacuum
values giving rise to corresponding domains and domain
walls. To understand the generic features of this
system of domain walls
one may consider the following idealised problem.
Imagine splitting space into cubes of the size of
correlation length. And, say, the probability for the
field to take a particular vacuum value in a given cube
is p ( $= {1 \over N }$, where N is the number of
degenerate vacuua ). The nature of domain structure
obtained, a domain being a set of connected cubes
carrying same vacuum value of the field, is a basic
question in percolation theory [7]. The main result we
will need here is that if p is greater than a certain
critical value ${p_c}$, then apart from finite size
domains there will be one and only one
domain of "infinite" size formed. For p less than
${p_c}$ there will not be any infinite size domain.
 Generically, i.e. considering even other lattices
than just cubic, ${p_c}$ happens to be less than 0.5 .
In the example of real scalar field we have been
considering, p equals 0.5 .
Thus there would be an "infinite" domain
corresponding to each vacuum and therefore infinite
wall of a very complicated topology.
Of course, there will also be a network of finite
size walls.
The question one is interested
in is the energy density contribution of this domain
wall system as it evolves. Following
crude analysis addresses this question [6] .

     The dynamics of the wall is mainly decided by the
force per unit area ${f_T}$ due to tension and
frictional force $f_F$ with surrounding medium. Since
tension in the wall is proportional to the energy
per unit area ${\sigma}$, we get
${f_T \sim}$ ${\sigma \over R}$ for radius of
curvature scale R. Moreover, $f_F {\sim sT^4}$ where
s is the speed of the wall and T the temperature
of the system [8]. When the speed of the wall
has stabilised we have

\begin{equation}
sT^4 = {\sigma \over R}
\end{equation}

\noindent Thus, the typical
time $t_R \sim {R \over s}$
taken by a wall portion of radius scale R
to straighten out would be

\begin{equation}
t_R \sim  {R^2T^4 \over \sigma} \approx {R^2 \over {G \sigma t^2}}
\end{equation}

\noindent Making the plausible assumption that
if ${t_R < t }$, the wall curvature on the
scale R would be smoothened out by time t, we
get that the scale on which wall is smooth
grows as

\begin{equation}
R(t) \approx (G \sigma)^{1\over 2} t^{3 \over 2}
\end{equation}

\noindent Energy density contribution $\rho_W$
to the universe by walls goes as

\begin{equation}
{\rho_W} \sim {{\sigma R^2} \over R^3 } \sim ({\sigma \over
{Gt^3}})^{1\over 2} \end{equation}

\noindent Therefore ${\rho_W}$ becomes
comparable to the energy density ${\rho} \sim {1\over {Gt^2}}$ of the
universe in the radiation dominated
era around
${t_0} \sim {1 \over {G\sigma}}$. Thus domain walls
would significantly alter the evolution of the
universe after ${t_0}$.

     Now, the discrete symmetries relevant
 for particle physics
typically tend to be broken at mass scales above
the weak scale ${M_w \approx 100 Gev }$, giving
$t_0 \leq {10^8}$ sec.. This would be
certainly true
of P and T (CP), the examples we are most
interested
in. Hence from above considerations one
would conclude
that discrete symmetries cannot be broken
spontaneously.

      There are two possible ways out of this impasse.
One possibility is that, even for low scales of symmetry
breaking, the phase transition that would have restored the symmetry
does not
take place, at least not until high enough temperatures
to allow inflation to dilute the energy density in the
domain walls. This in general requires a more complicated
Higgs structure than the minimal one
and realistic examples have been discussed in the
literature [10].

 Another way out [6] , the one of interest
to us in this paper, is the possibility that a spontaneously
broken discrete
 symmetry is also explicitly broken by a small
amount, which lifts the degeneracy of the two vacua
$+v$ and $-v$. For instance, in our example we could imagine
adding to the Lagrangian a small ${\phi}^3$
term which, obviously,
breaks ${\phi} \rightarrow -{\phi}$ symmetry.
 It should not come as a
surprise that this effect may provide a mechanism for
the decay of domain walls; after all now there is a
unique vacuum. Crudely, the way it works is as
follows [6]. Lifting of the degeneracy of the two
vacua by an amount ${\epsilon}$ gives a pressure
difference of the same amount, between
the two sides of the wall, with
a tendency to push the wall into false vacuum region.
Thus the dynamics of the wall is now going to be
decided by combination of the pressure ${\epsilon}$
, forces $f_T$ due to
tension and $f_F$ due to friction mentioned before.
Clearly at some point the forces due to friction and
tension become small, compared to pressure difference
${\epsilon}$, because they are
proportional to $T^4 \sim {1\over Gt^2}$ and
${\sigma \over R} \sim ({\sigma \over {Gt^3}})^{1\over 2}$
respectively. At that
stage the pressure difference will dominate and
cause shrinking of the false vacuum. Actually it is
difficult to find out precisely when the false
vacuum region, and hence the domain walls, disappear.
However, it may be crudely estimated to be the time
when the pressure ${\epsilon}$ exceeds
the force due to tension, or when it exceeds the force
due to friction for a relativistically moving
wall so as to dominate the dynamics. For either
requirement to be satisfied before
$t_0 \sim { 1 \over {G \sigma} } $, the time
when wall contribution $\rho_W$ would
have become comparable to the energy density
of the universe, one obtains

\begin{equation}
\epsilon \geq G{\sigma}^2  \sim  { {v^6} \over {{M_{Pl}}^2}}
\end{equation}

      Of course, it is not very
appealing to introduce ad hoc
the symmetry breaking terms just in order to eliminate
the domain-wall problem. Ideally,
we would prefer these effects
 to be a natural consequence of
underlying theory. An interesting example recently
discussed in the literature
[11] is that of a discrete symmetry
explicitly broken due to instanton induced effects.

{\bf Role of Gravity}

     In this paper we invoke the possibility that the needed
mechanism for explicit breaking
may be naturally provided by gravity. One
expects that gravity, because of black-hole physics,
may not respect global symmetries, both continuous
and discrete ones.
This expectation is motivated by two important points:
 firstly, the
``no-hair" theorems of black-hole physics that state that
stationary black-holes are completely characterised by quantum
numbers
associated with long-range gauge fields, and secondly,
that the Hawking radiation in evaporation of
black-hole is thermal [12]. Now, consider a process in
which a certain amount of normal matter, which is in a
state that is ``odd" under the discrete symmetry in
consideration, collapses under gravity to form a
black-hole. Because of no hair being associated to
the global discrete symmetry, any information
regarding it is lost to observers outside the black
hole. Hawking radiation from the black hole
being thermal in nature does not carry any
information about internal states of the black-hole
either. Of course, it is not certain what the
properties of evaporation would be at late stages
when semi-classical approximation breaks down.
Unless for some reason the processes at late stages
cause the final system to have same global
discrete charges as those of the initial normal
matter that collapsed, the symmetry would stand
violated.

     We wish to note that from very different
viewpoints there have been discussions in the
literature regarding possiblity of CP or T violation
in context of gravity. Ashtekar et. al. have
discussed [13] CP "problem" in the framework of
canonical quantisation of gravity.
In Ashtekar variables
reformulation of general relativity, the canonical
variables of the theory resemble those of Yang-Mills
theory. This allows for discussion of ${\theta}$
sectors in canonical quantisation framework for
Yang-Mills to be taken over to the gravity case.
 Moreover, an analogue
of ${\theta FF^d}$ term in the action can also be
given.

      Another set of observations that interest us
particularly were made by Penrose about
T-asymmetry [14]. He contends, based on arguments
related to Bekenstein-Hawking formula,
that there must be some as yet unknown theory
 of quantum gravity that is
time-asymmetric.
 We recall
here only an easy to state, interesting point from
his discussion.
Corresponding to a solution of Einstein's
equation describing collapse of normal matter to
form a black-hole that stays for ever (classically)
, there would be a time-reversed solution,
white-hole, describing
explosion of a singularity into normal matter.
Now, according to Bekenstein-Hawking formula
the surface area $A$, of a black hole's
horizon is proportional to its intrinsic
entropy, $S$

\begin{equation}
S = {kc^3 \over {4\hbar G}} A
\end{equation}

\noindent In classical processes area is
non-decreasing with time and hence the entropy.
If an intrinsic entropy is associated with a
white-hole, it is again expected to be proportional to
the area of its horizon. Time
reverse of area principle would give that this
area, and hence the corresponding entropy
can never increase, an anti-thermodynamic behaviour.
 Especially it would be
a strongly
anti-thermodynamic behaviour by the white hole
when it ejects substantial amount of matter.
This is among the reasons that lead Penrose to
consider the possibility that there may be a
general principle that rules out the existence
of white holes and would therefore be time-asymmetric.

     With the premise, in view of preceeding
 discussion, that gravity may violate a
global discrete symmetry we wish to explore
 its consequences for the domain-wall problem.

     The crucial issue one faces in
implementing
this kind of approach is determination of the
precise form of these symmetry-breaking terms. At the
present day understanding of gravity it does not seem
possible to give a satisfactory answer to this question.
The strategy followed in the literature [15], which we also
adopt here,
in analogous discussions
 has been to write
all the higher dimensional effective operators allowed
by gauge invariance of an underlying theory. Of course,
one could take a point of view that the dimension
four and lower terms may also break the discrete
symmetry. We take no stand on this point. In any
case, even if this happens it can only help in
destabilising the domain walls due to increased
symmetry breaking. Our point is that even the tiny
higher dimensional symmetry breaking terms,
cut-off by powers of Planck-mass, may be sufficient
in solving the domain wall problem.

  To illustrate how this works, we turn again to our
simple example of a real scalar field. The effective
higher dimensional operators would take the form

\begin{equation}
V_{eff} = {C_5 \over {M_{Pl}}}{\phi}^5
+ {C_6 \over {M_{Pl}^2}}{\phi}^6 + ....
\end{equation}

     Obviously, all the terms with odd powers of
 ${\phi}$ break the discrete symmetry ${\phi} \rightarrow -{\phi}$.
[We should mention that
while discussing difficulties with a certain
compactification scheme in superstring theory, it
was remarked by Ellis et. al. [16] that a specific
discrete symmetry in their model may be broken by
terms inversely proportional to ${M_{Pl}}$.
But they note further
that massless modes of string theory
would not induce such
terms and that the massive modes, could be the
only possible source of such effects. However, as
we have been pursuing here, the non-perturbative
effects are expected
to be a natural source of breaking
of global discrete symmetries, independent of whether
string theory turns out to be a correct theory of
gravity. Furthermore, as we have emphasised before,
we feel that these effects should be taken
seriously as a possible solution to the domain-wall
problem associated with the fundamental discrete
symmetries of nature such as Parity or Time-reversal
invariance.]

     In estimating precisely the amount of
symmetry-breaking we would need to know the values
of coefficients $C_n$. Barring some
unexpected conspiracy, in the following we will assume
that $C_n$ will be O(1) as they are dimensionless. Moreover,
it is understood that the scale of spontaneous
symmetry-breaking $v$ lies below the Planck scale.
With all this in mind, the
energy-density split between the two vacuua would be
$\approx {C_5 \over M_{Pl}}v^5$.
 This is obviously much bigger than the amount
 $v^6 \over {M_{Pl}^2}$ needed to make the domain walls disappear.
This holds true as long as $C_5 \gg {v \over {M_{Pl}}}$,
which for lower scales of symmetry breaking gets
to be more and more plausible. For example,
if $v = M_{GUT} \approx 10^{15}$ Gev, we need $C_5 > 10^{-4}$,
 whereas for
$v = M_w \approx 100$ Gev, we only need $C_5 > 10^{-17}$ !
 In general, if a leading operator in eq.(10) is of dimension n,
the condition for disappearance of domain walls is
 $C_n > ({M_{Pl} \over v})^{n - 6}$. Clearly $n=6$ is the critical
value, since for $n>6~C_n$ would have to be unreasonably large
whereas for $n=6~C_6{\sim}O(1)$ may suffice.

{\bf Examples of $T$ and $P$ }

      An important issue that remains to be discussed
is what happens in realistic examples of discrete
symmetries associated with gauge theories of
strong and electroweak interactions. This clearly
must be case dependent. Here we study two
discrete symmetries of central importance : CP (T) and P.

  $CP$ or $T$ invarince: The simplest and most popular
example of the spontaneous breaking of CP is the
 two Higgs doublet version of the standard model.
One simply imagines all the couplings in the
Lagrangian to be real and looks for minimum of
energy with non-trivial phases. By an $SU(2){\times}U(1)$
gauge transformation we can write the most general
solution in the form

\begin{equation}
< {\phi}_1 >~=~ \left(
\begin{array}{c}
       0 \\
       v_1
\end{array}
\right)~,~
< {\phi}_2 >~=~ \left(
\begin{array}{c}
       0 \\
       v_2e^{i\delta}
\end{array}
\right)
\end{equation}

\noindent where ${v_i, \delta}$ are real
 numbers. Through the
terms of the type ${\phi}_1^{\dagger}{\phi}_2$
and $({\phi}_1^{\dagger}{\phi}_2)^2$ in the potential
, the theory knows of the phase ${\delta}$ and in
general we can write

\begin{equation}
{V = A + B cos\delta + C cos2\delta}
\end{equation}

 For a range of parameters there is a solution
with non-vanishing and non-trivial phase ${\delta}$,
giving spontaneous breaking of CP.
Hence the domain wall problem.

 If we now follow our logic of expanding the
effective higher degree potential induced by gravity
in inverse powers of $M_{pl}$, keeping gauge invariance intact,
the leading term is of dimension six

\begin{equation}
{\Delta}V = {c_{ij} \over {M_{Pl}^2}} ({\phi}_i^{\dagger}{\phi}_j)^3 + h.c.
+ ~higher~order~terms
\end{equation}

 Again, the possibility of
 breaking of CP by gravity
would be reflected in complex coefficients
$c_{ij}$. As we observed before, only for the dimensionless
coefficient $C_6 \geq O(1)$
the domain walls would be unstable.
 Thus, the understanding of detailed consequences
of gravity turns out to be
crucial in such a low enegy issue as CP
violation and we believe that any hint in
this direction is extremely important.
  The situation would change dramatically
if one is willing to introduce a complex
singlet into the theory and attribute to
its complex
vacuum expectation value the source of
spontaneous CP violation. Now clearly
the leading operator, analogous to our simple
example of a real scalar field, is of
dimension five.

     $Parity:$ The simplest models which describe
spontaneos breaking of Parity are based on
$SU(2)_L \times SU(2)_R \times U(1)$
gauge group with $g_L = g_R$ gauge
couplings [17]. The breaking of Parity is
attributed to the large mass for the
right-handed gauge boson: $M_{W_R} \gg M_{W_L}$. By
introducing Higgs multiplets ${\phi}_L$ and ${\phi}_R$
which are nontrivial representations
( doublets or triplets ) under $SU(2)_L$
and $SU(2)_R$ respectively , the above can
be achieved through spontaneus breaking of
Parity: $|< {\phi}_R >| \gg |< {\phi}_L >|$
  ( where $P: {\phi}_L {\leftrightarrow} {\phi}_R$ ) .
 The analysis of gravitational effects
parallels completely our discussion on
CP. If P is broken through the $v.e.v.$ of
nonsinglet then the leading P breaking
operators must be of dimension six or larger,
and the fate of domain walls will depend
on the dimensionless parameters which
characterise breaking terms. Another
possibility is the existence of parity
odd singlet ${\sigma}$ ( $P: \sigma {\rightarrow} -{\sigma})$,
 and P being
broken through $< \sigma > \neq 0$ . In this case, the
leading P breaking gravity induced term could be
of dimension five.

{\bf Conclusion}

      In this paper we have
taken seriously the possibility
that gravity breaks global discrete
symmetries. If so, this could be a natural
source of instability of domain walls. We find it
curious that such a mechanism to solve the domain
wall problem does not involve any other scales than
$M_{Pl}$ and the scale of spontaneous
symmetry breaking $v$, already present in the theory.
Although our main examples were of CP (T) and P,
by no means we wish
to imply that the effects of gravity have to
stop there. As it emerged from our
discussion, whether gravity can play the
desired role depends on the lowest degree
induced effective operator.

      We are well aware of the speculative nature of
 our suggestion so that one cannot yet be certain that
it is actually realised. However, we hope to have conveyed
the necessity of further understanding of gravity before
one can claim the existence of the domain wall problem.

  {\bf Acknowledgement}

    It is a pleasure to thank G. Dvali, R. Mohapatra and V. Rubakov
for discussions and encouragement.

\end{document}